# Forecasting Crime Using ARIMA Model

**Khawar Islam, Akhter Raza**

[1]*Computer Science, Federal Urdu University, University Road Karachi, Karachi, Sindh, Pakistan, E-mail: khawarislam@fuuast.edu.pk*

**Abstract**
Data mining is the process in which we extract the different patterns and useful Information from large dataset. According to London police, crimes are immediately increases from beginning of 2017 in different borough of London. No useful information is avail- able for prevent crime on future basis. We forecasts crime rates in London borough by extracting large dataset of crime in London and predicted number of crimes in future. We used time series ARIMA model for forecasting crimes in London. By giving 5 years of data to ARIMA model forecasting 2 years crime data. Comparatively, with exponential smoothing ARIMA model has higher fitting values. A real dataset of crimes reported by London police collected from its website and other resources. Our main concept is divided into four parts. Data extraction (DE), data processing (DP) of unstructured data, visualizing model in IBM SPSS. DE extracts crime data from web sources during 2012 for the 2016 year. DP integrates and reduces data and give them predefined attributes. Crime prediction is analyzed by applying some calculation, calculated their moving aver- age, difference, and auto-regression. Forecasted Model gives 80% correct values, which is formed to be an accurate model. This work helps for London police in decision-making against crime.

## 1 Introduction

Crime is an activity or unpredictable scene against society. The increase of population directly effects on country resources where government responsibility is to manage re- sources allocating resources on right place. Crime is an activity which faces all developing and developed countries [1]. Many techniques are applied to analyze crime patterns and identify places where chances of crimes are maximum in the future. Some crimes applications are implemented and followed by the government [10]-[11]-[20]. Here, we consider and dis-cuss the data mining technique for prediction with the availability of different datasets, many researchers and experts used this datasets foe future prediction. Many predictions are may be accurate or not depend upon the situation. Crime prediction has always remained a hot topic in data mining, because law enforcement agencies used these predictions and takes different steps.

---

*Corresponding author.
†E-mail: khawarislam@fuuast.edu.pk

In the modern era, where researchers implemented many techniques to minimize or identifies crime patterns and identification using analysis of historical crime data and their trends. In result, crimes are minimized, but we can't get rid of crimes. One of the most famous crime attacked the world trade center on 11, 2001. Some popular crimes in London are listed down according to their occurrence. Torso human floating in River Thomas on 2001 [2]. Murder of Sally Anne Bowman who worked as a model and hairdresser was raped in 2015 [3]. Ben Kinsella was an English student, he's murdered by a gang of black teenagers in 29, June 2008 [4]. Tia Sharp was a high-profile case of child murder her dead body found in her grandfather in August 2012 [5]. Gemma Mccluskie who worked as an actress, her body discovered in the Regent Canal on March 2012 [6]. Lee Rigby walked on Wellington Street, two men killed him with knives on May 2013 [7]. Alice was a 14-year-old girl, missing August 2014 and found her body in Boston Manor Park on 28 August 2014 [8]. Many crimes became the headline of television. Law enforcement agencies, researchers, and computer analysts are working together for many years to minimize crime rate.

We proposed an approach to the forecasting crime rate in a different borough of London and forecast crime pattern graph using historical data about crimes and gives crime trend for future years. Our nature of a problem is to predict crime rate in London for future by the availability of the dataset. The purpose of this research study is to highlight those boroughs of London, where crime rate will be increase according to time frame. This research is helpful to law enforcement agencies to work out in limited resources. As no previous work done using ARIMA model to forecast crime rate.

## 2   Methods

Our section is divided into three subsections: Sect 3.1, 3.2 and 3.3. Section 3.1 describes London crime (LC) dataset for choosing reasons. Section 3.2 describes the LC selection because LC data set contains data of all boroughs of London with appropriate resources. Section 3 describes data mining technique which we are applying for crime forecasting.

### 2.1   LC Dataset

Dataset gathering is challenging part of crime analyses and prediction. Crime dataset is taken from thirty-four borough of London. The reason of the selection of borough of London, because London police maintain rich data of crime on a daily basis and distributed in monthly wise. Most of the verified data collected from their verified sources. All the crimes are distributed in month wise of each city.

## 2.2   LC selection of London dataset

The LC dataset approaches to choose four boroughs of London (Barking and Dagenham, Barnet, Bexley and Brent) based upon crime rates in this borough. A total number of crimes from 2012 to 2016 are analyzed using graph (please refer Fig 1) shows the borough wise distribution of crime in each borough. In line graph y axis shows a total number of crimes in each city and x-axis show year wise distribution.

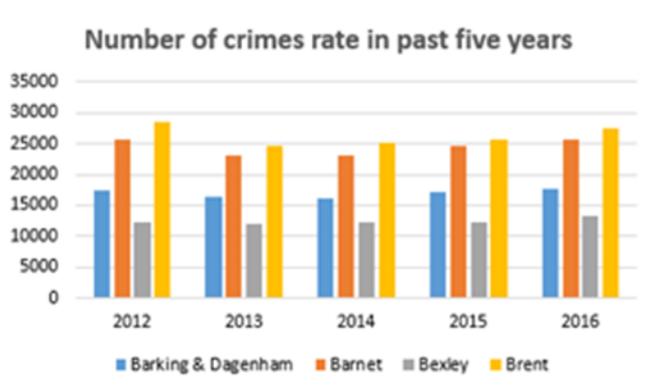

Fig 1. Number of crimes versus year for four London boroughs

## 2.3   Forecasting Technique

We start from LC dataset using two ways (1) collected crime records in unstructured form from a web source, namely – Data police UK [7]. Data available from web sources during the period 2012 – 2016. (2) Applying data processing techniques to clean unstructured crime data and extracted into structured data with 2100 crime instances (.csv format). The structured crime data represented borough and monthly crime rate attributes. The structured data is implemented using two tools. (1) Microsoft Excel (2) IBM SPSS. Microsoft Excel for data cleaning and processing. IBM SPSS for crime prediction. Figure 2 shows LC dataset.

Fig 2. Structured data of London crime data set taken from the UK police website [47]

## 3   Experiments

### 3.1   LC Dataset Approach

In this section, we discussed the flow of the proposed LC dataset using Fig 3. The LC data set consists of (.csv) format. LC dataset is fully integrated with various techniques of data mining, including data cleaning, integration, and reduction. IT gains us more flexibility to detect crime patterns and predicted values. LC dataset will help law enforcement agencies to predict crime of different borough and its rate. The proposed work flow starts from LC dataset. We find and search London crime data from different websites and other sources, then we follow data cleaning process to clear data and remove raw data (Missing value). The LC dataset supplied with IBM SPSS software to critical analyses of crime data, we perform some experiment through different model included regression. Linear regression and ARIMA model. Finalized Model is an ARIMA model because the performance of ARIMA model is more accurate and forecasting values are comparatively similar to crime happened. On the basis of prediction, law enforcement agencies manage police, according to the crime rate in each city.

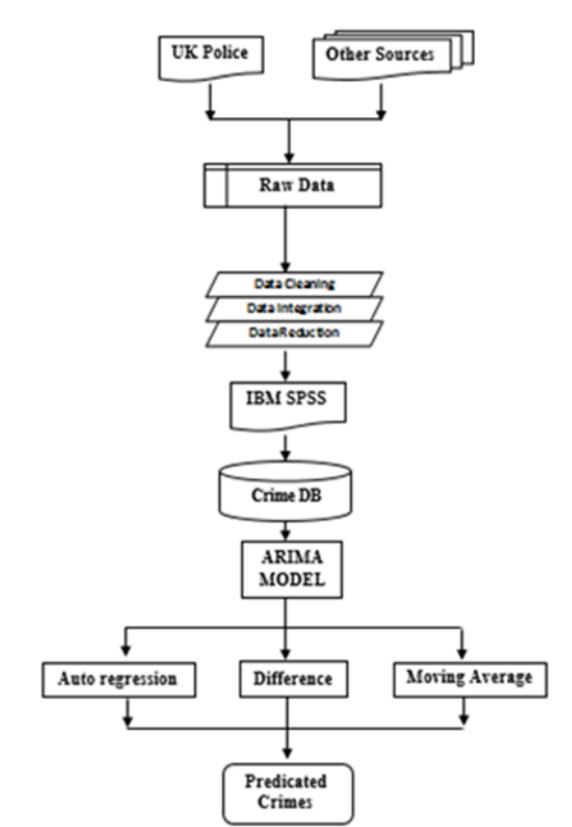

Fig 3. London Crime prediction work flow

### 3.2   Model & Algorithm Selection

Different algorithms are designed to analyze and identify a pattern of data in which machine learning algorithms are used to predict values like linear regression, Decision tree, Support vector machine, Random Forest etc. Many data science algorithms are used in determining the data pattern like visualization, text mining, and auto regression moving the average model. The ARIMA model is selected to conduct an analysis of our dataset.

Forecasting economic, industry, financial, marketing, population purposes ARIMA model is a suitable selection across different models. Auto regression integrated moving average (ARIMA) is successfully used for prediction. This model was created in 1970 by Gwilym, M.Jenkins and George E.P Box.

The forecasting ARIMA model of stationary time series of our crime dataset is

The predicted value of the crime data set is the = Constant / Sum of one or more recent values of Y and recent values of error of Y

In crime model, Stationaries series called auto regression, forecasting of errors called moving average and the series, which is different to be made stationary called integrated. ARIMA model is constructed with (p, d, q) where p is auto regression, d is the nonseasonal difference, q is logged forecast error. The forecasting equation of crime data is constructed Y, denoted with a depth difference of Y.

If d = 0

$$Y_t = Y_t$$

If d = 1

$$Y_t = Y_t - Y_{t-1}$$

If d = 2

$$Y_t = (Y_t - Y_{t-1}) - (Y_{t-1} - Y_{t-2})$$

The forecasting equation is

$$\hat{Y}_t = \mu + \phi_1 y_{t-1} + \ldots + \phi_p y_{t-p} - \theta_1 e_{t-1} - \ldots - \theta_q e_{t-q}$$

Where moving average is defined by (θ's) in terms of positive and negative sign. For crime dataset, we first estimate different values in auto regression model. The best fit value of crime data is

ARIMA (2, 0, 0)

The forecasting equation is

$$\hat{Y}_t = \mu + \phi_1 Y_{t-1}$$

Now, we find series Y, crime data is not stationary, so predicted values are

ARIMA (0, 2, 0)

The general forecasting equation is

$$\hat{Y}_t - Y_{t-2} = \mu$$

$$\hat{Y}_t = \mu + Y_{t-2}$$

Both values are correlated with each other. We can add a dependent variable for forecasting

$$\hat{Y}_t - Y_{t-2} = \mu + \phi_1(Y_{t-1} - Y_{t-2})$$

$$\hat{Y}_t - Y_{t-1} = \mu$$

The generalized equation of forecasting is

$$\hat{Y}_t = \mu + Y_{t-1} + \phi_1 (Y_{t-1} - Y_{t-2})$$

The both values provide confidence level 1 which means that the interval will be accepted.

LC dataset detects crime patterns and relation between crime data using ARIMA model techniques. These techniques provide us to identify crime and facility in handling crime information in each city. This data set is used for crime prevention. Crime based on each city helps law enforcement agencies and government to take proper measure of police against a criminal. For example, Camden has the highest rate of crime report among seven boroughs of London. So law enforcement agencies should arrange special place or increase police for this city.

## 4  Results

In this section, we provide implementation details for the proposed approach which gives a better result for long term crime forecasting. The outcomes, values and graph of models are shown in Fig 4, 5, 6 and 7. As in our pictures, it is found that the model fits with its series and predict better values which are also useful in a real scenario. The overall accuracy of the ARIMA algorithm is best for crime prediction. The algorithm gives the best value of R-squared among lowest error value. To verify the model, we take from 2012 to 2015 and predict 2016, the actual and predicted value is very close to reality. Because of the accurate forecasting of data, we predicted crime rates for 2017 to 2018.

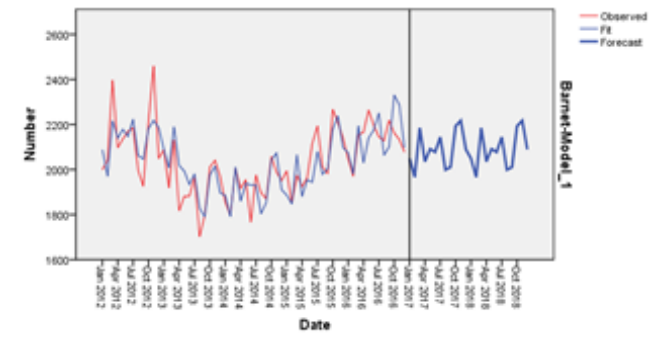

Fig 4. Forecasting Model of Brent Borough in London

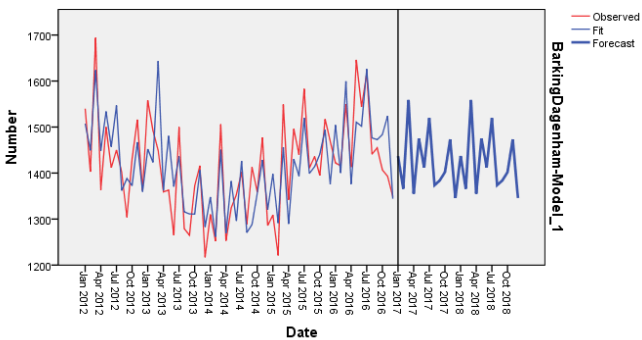

Fig 5. Forecasting Model of Barking Borough in London

This result is beneficial and important for crime suppression for the local government and police stations. Because of the accurate predictions of crimes of the ARIMA model. We will take some future emergency measurements, such as criminal activities, patrolling will be patrolling will be prepared in advance and limited resources will be deployed rightly to minimize the crime rate decision making will be improved greatly for the local police station and municipal governments.

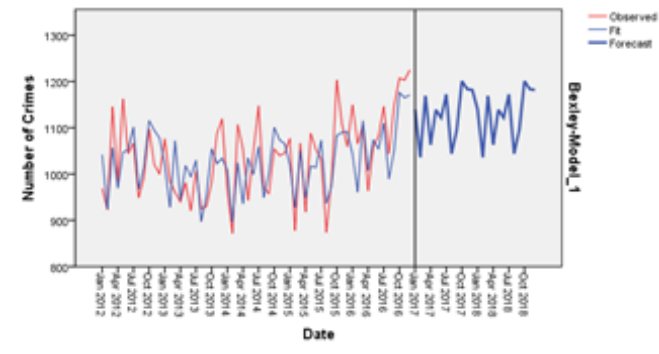

Fig 6. Forecasting Model of Bexley Borough in London

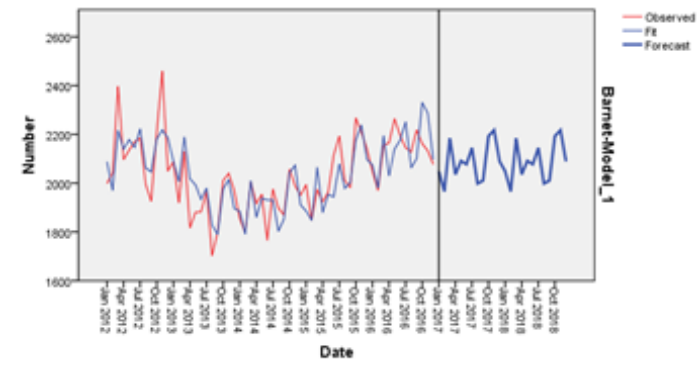

Fig 6. Forecasting Model of Barnet Borough in London

## 5   Conclusions

Crime in London borough is continuously going on, which is dependent upon some factors such as poverty, unemployment, frustration, etc. Police investigation agencies take an active role against criminal but it takes much more time. So we contribute towards for forecasting crime rate in a different borough of London. Our proposed model extracts unstructured data from web sources including London police website. Investigating agencies should use our proposed model to check the crime rate in the future. This model helps and speeds up a process where crime rate will become higher. We used time series ARIMA model is used to calculate crime rate for 2 years (2017 to 2018) by providing five years past data (2012 to 2016). This study presented the best fit model for crime rates in London. The comparison between past crime rate and forecast crime rate behaved well and this statistical record is also used in real scenarios. The table 2 presents the forecasting values of 2017 and 2018. For predicting purpose, we take values of present year 2017 for matching predicted values and actual values which is 80% correct. Table 2 shows number of crime according to month. The upper confidence limit (UCL) shows higher chances of crime. The lower confidence limit shows lowest chances of crime. The forecast value is predicated value of crime based on data.

TABLE 1: MODEL STATISTICS

| Model | Number of Predictors | Model Fit statistics | | Ljung-Box Q(18) | | | Number of Outliers |
|---|---|---|---|---|---|---|---|
| | | Stationary R-squared | R-squared | Statistics | DF | Sig. | |
| Barking Dagenham Model | 0 | .676 | .588 | 26.405 | 16 | .049 | 0 |
| Barnet-Model | 0 | .636 | .636 | 16.278 | 16 | .434 | 0 |
| Bexley-Model | 0 | .743 | .535 | 38.989 | 16 | .001 | 0 |
| Brent-Model | 0 | .614 | .690 | 4.971 | 16 | .996 | 0 |

TABLE 2: MODEL STATISTICS

| Model | Barnet-Model | | | Barking Dagenham-Model | | | Bexley-Model | | | Brent-Model_1 | | |
|---|---|---|---|---|---|---|---|---|---|---|---|---|
| | Forecast | UCL | LCL | Forecast | UCL | LCL | Forecast | UCL | LCL | Forecast | UCL | LCL |
| Jan-17 | 2048 | 2231 | 1866 | 1437 | 1574 | 1300 | 1139 | 1260 | 1019 | 2363 | 2567 | 2159 |
| Feb-17 | 1967 | 2164 | 1770 | 1366 | 1513 | 1219 | 1037 | 1160 | 913 | 2191 | 2430 | 1953 |
| Mar-17 | 2184 | 2394 | 1973 | 1559 | 1716 | 1402 | 1169 | 1294 | 1043 | 2373 | 2641 | 2105 |
| Apr-17 | 2036 | 2258 | 1813 | 1355 | 1522 | 1189 | 1064 | 1192 | 936 | 2204 | 2498 | 1909 |
| May-17 | 2091 | 2325 | 1857 | 1475 | 1650 | 1300 | 1139 | 1269 | 1008 | 2416 | 2736 | 2097 |
| Jun-17 | 2077 | 2323 | 1832 | 1412 | 1596 | 1228 | 1122 | 1254 | 989 | 2358 | 2700 | 2016 |
| Jul-17 | 2144 | 2400 | 1888 | 1520 | 1711 | 1328 | 1172 | 1306 | 1037 | 2358 | 2721 | 1995 |
| Aug-17 | 1999 | 2265 | 1732 | 1374 | 1573 | 1174 | 1044 | 1181 | 908 | 2285 | 2669 | 1902 |
| Sep-17 | 2010 | 2286 | 1734 | 1384 | 1590 | 1177 | 1095 | 1234 | 956 | 2230 | 2633 | 1828 |
| Oct-17 | 2193 | 2479 | 1907 | 1402 | 1616 | 1188 | 1201 | 1342 | 1060 | 2440 | 2861 | 2020 |
| Nov-17 | 2217 | 2512 | 1922 | 1473 | 1694 | 1252 | 1183 | 1326 | 1040 | 2422 | 2860 | 1984 |
| Dec-17 | 2089 | 2393 | 1785 | 1346 | 1573 | 1119 | 1182 | 1327 | 1036 | 2277 | 2732 | 1822 |
| Jan-18 | 2048 | 2361 | 1736 | 1437 | 1671 | 1203 | 1139 | 1287 | 992 | 2363 | 2834 | 1892 |
| Feb-18 | 1967 | 2288 | 1646 | 1366 | 1606 | 1126 | 1037 | 1186 | 887 | 2191 | 2678 | 1704 |
| Mar-18 | 2184 | 2513 | 1854 | 1559 | 1805 | 1313 | 1169 | 1320 | 1018 | 2373 | 2875 | 1871 |
| Apr-18 | 2036 | 2373 | 1698 | 1355 | 1608 | 1103 | 1064 | 1217 | 911 | 2204 | 2721 | 1687 |
| May-18 | 2091 | 2436 | 1746 | 1475 | 1733 | 1217 | 1139 | 1293 | 984 | 2416 | 2948 | 1885 |
| Jun-18 | 2077 | 2430 | 1725 | 1412 | 1676 | 1148 | 1122 | 1278 | 965 | 2358 | 2903 | 1813 |
| Jul-18 | 2144 | 2504 | 1784 | 1520 | 1789 | 1250 | 1172 | 1330 | 1013 | 2358 | 2917 | 1799 |
| Aug-18 | 1999 | 2366 | 1631 | 1374 | 1649 | 1099 | 1044 | 1205 | 884 | 2285 | 2857 | 1713 |
| Sep-18 | 2010 | 2385 | 1635 | 1384 | 1664 | 1103 | 1095 | 1257 | 933 | 2230 | 2816 | 1645 |
| Oct-18 | 2193 | 2575 | 1811 | 1402 | 1688 | 1116 | 1201 | 1365 | 1037 | 2440 | 3038 | 1842 |
| Nov-18 | 2217 | 2606 | 1828 | 1473 | 1764 | 1182 | 1183 | 1349 | 1018 | 2422 | 3032 | 1812 |
| Dec-18 | 2089 | 2485 | 1693 | 1346 | 1642 | 1050 | 1182 | 1349 | 1014 | 2277 | 2900 | 1655 |

## 6   Discussion

Literature survey includes many approaches on crime prediction with some limitation. Different authors discussed different approaches for crime prediction. Most of the common discussion is crime classification using different clustering algorithms. Aghababaei et al. [9] discussed crime prediction based on posted tweets, however, they don't provide training dataset to evaluate results and measure performance accurately. Although some authors S.R Deskmukh et al. 2015; Akshay Kumar Singh et al. 2016; A.Bharathi et al. 2014; Tushar Sonaqwanev et al. 2015; [10]-[11]-[12]-[13] discussed crime prediction using data mining techniques, apply different algorithms K-Mean, Apriori algorithms, naive Bayes classifiers. Thongtae and Sirsuk 2008; [14] Shiju Sathyadeven et al. 2014; [15] Lawrence and Natarajan 2015 [16]; Malathi et al. 2011; [17] discussed crime classification, patterns based on the training dataset. Training data set used as knowledge discovery for future predictions. Although some work is done in crime detection Shyam Varan Wath 2006; [18] Chung Hisen et al. 2011; [19] Vineet et al. 2016 [20] discussed semi-supervised learning, support vector machine (SVM), a decision tree for crime detection. Tahani et al. 2015; [21] discussed crime prediction based upon the spatial and temporal dataset, compare two data sets and identify crime types. S.Sivaranjani et al 2016; [22] work with crime activates using density based spatial clustering and different algorithms. Pen Chang et al 2008; [23] discussed an ARIMA model for forecasting crimes. Zakaria Suliman and Ayman Altaher 2013; [24] Anisha Agarwal et al. 2016; [25] Anshu and Raman Kumar et al. 2013; [26] proposed a model through crime data, analyze and suggest predicted the desired pattern. Lenin Mookiah et al. 2015; [27] conducted crime survey in which they collected different crime variables and rate. Tirthraj and Rajanikanth 2016; [28] Sushant Bharti and Ashutosh 2015; [29] developed a predictive model for crime analytics, which helps law enforcement authorities to allocate resources to higher areas where crime chances are maximized. Although Abdul Awar et al. 2016; [30] discussed same work, but using linear regression to predict future crime. Some work discussed on crime classification Addarsh, Abhilash, and Poorna; [31]. Umair Saeed et al. 2015; [32] have analyzed crime patterns by applying machine learning algorithms. Using an available dataset on internet web based work also discussed Xinyn Chen et al 2015; [33] discussed crime prediction using a twitter dataset to predict crime and location. Mark B. Mithchell et al 2007; [34] introduced web based crime toolkit to analyze crime. They developed a tool for forecasting crime in Richmond city. Suzilah and Nurulhuda 2013; [35] identify crimes based on forecasting technique. Adjusted decomposition techniques are used to detect crimes. The research was taken out of Kedah city located in Malaysia. Concluding remarks show that crime index is higher in the future, which directly effect on the economy of Malaysia. Gabriel Rosser et al. 2016; [36] discussed network based crime mapping, a calibrated based model outperformed than grid alternative. 20% crimes are more identified through this model. Aziz Nasiridinov et al. 2014; [37] detect crime patterns using historical dataset. Four steps are followed to predict crimes. 1) Algorithm selection 2) Generation steps 3) Result 4) Performance accuracy. The proposed system is given to law enforcement agencies to detect or identify crimes. Tong Wang et al. 2017; [38] Discussed and analyze different learning patterns for crimes. A robustness are automatically detected crime pattern on a large dataset to save time-consuming process. The purpose of the crime pattern detection algorithm known as Series Finder. Andrey Bogomolvo et al. 2014; [39] presented novel research on crime based mobile data. Analyze human behavior from mobile to make crime prediction data of human with basic geographic information. They obtain 70% accurate result of predicted crimes. Bruno Cavadas et al. 2015; [40] proposed linear regression technique applied to violet crime dataset. Learning system gives the best performance on prediction. This system is limited to the USA. Vikas Grover et al. 2006; [41] examined different techniques of crime prediction, including statistical method, offending behavior, and geographical information. Many researchers spent a lot of time on data analyzing. Limited techniques are available. Anchal Rani and Rajasree 2014; [42] analysis of crime trends, Mahanolobis, Euclidean and Minkowski distance model and time warping technique are used in multivariate time series to identify crimes. Wilpen et al. 2013; [43] forecasting crime by selection of one month ahead and apply univariate time series with the naïve method. The predicted values are more accurate than police practices. Michael Hanslamaier et al. 2015 [44] forecasting crime in Germany, model forecasting values till 2020. Based on their research offenses are expected to increase in future. Hanmant et al. [45] Forecasting short term crime rate in Satara district. Secular trend analysis is used to forecast crime value in short time. Saoumya and Baghel 2015; [46] proposed a predictive model using big data. Crime mapping algorithm identifies that area which is highly affected. Valuable data used in Artificial Neural Network for future research on crime trend.

**References**

1. Crime. (2018). https://en.wikipedia.org/wiki/crime (accessed on August 14, 2017).

2. A murdered African boy whose torso was found in the River Thames in 2001 and whose identity has remained a mystery has been named by a key witness. (2013). https://theukdatabase.com/cold-cases-missing-murdered-uk-kids-can-u-help/keith-lyon-sussex-1967/torso- case-boy-river-thames-2001 (accessed on August 14, 2017).

3. Heartbroken mum of murdered Sally Anne Bowman shares pain as killer reveals details of her daughter's horrific death. (2018). https://www.thesun.co.uk/news/4129605/heartbroken-mum-of-murdered-sally-anne- bowman-shares-pain-as-killer-reveals-details-of-her-daughters-horrific-death/(accessed on August 18, 2018).

4. Ben Kinsella murder: how a misplaced glance led to innocent teen's murder. (2009). http://www.telegraph.co.uk/news/uknews/law-and-order/5493572/Ben-Kinsella-murder-how-a-misplaced- glance-led-to-innocent-teens-murder.html (accessed on August 11, 2018).

5. Tia Sharp death: Stuart Hazell accused of killing her after sexual assault. (2013). https://www.theguardian. com/uk/2013/may/07/tia-sharp-death-sexual-assault (accessed on August 14, 2018).

6. The Murderer of East Enders actress Gemma McCluskie in line to pocket NHS compensation payout. (2014). https://www.mirror.co.uk/news/uk-news/murderer-eastenders-actress-gemma-mccluskie-4096560 (accessed on August 18, 2018)

7. Murder of Lee Rigby verdict. http://www.bbc.com/news/uk-22644057 . (2014). (accessed on 18 August)

8. Alice Gross murder: police say enough evidence to have charged Arnis Zalkalns.(2014). https://www.theguardian.com/uk-news/2015/jan/27/alice-gross-murder-arnis-zalkalns . (accessed on August 18, 2017)

9. Aghababaei, Somayyeh, and Masoud Makrehchi. "Mining social media content for crime prediction." 2016 IEEE/WIC/ACM International Conference on Web Intelligence (WI). IEEE, 2016.

10. S.R.Deshmukh, Arun S. Dalvi, Tushar J .Bhalerao, Ajinkya A. Dahale , Rahul S. Bharati, Chaitali R. Kadam, Crime Investigation using Data Mining. International Journal of Advanced Research in Computer and Communication Engineering Vol. 4, Issue 3, March 2015

11. Akshay Kumar, Neha Prasad, Nohil Narkhede, and Siddharth Mehta. "Crime: Classification and pattern prediction." International Advanced Research Journal in Science, Engineering and Technology 3, no. 2 (2016): 41-43.

12. Bharathi, A. Shilpa, and R. Shilpa. "A survey on crime data analysis of data mining using clustering techniques." International Journal of Advance Research in Computer Science and Management Studies 2.8 (2014): 9-13.

13. Sonaqwanev, Tushar, Shirin Shaikh, Shaista Shaikh, Rahul Shinde, and Asif Sayyad. "Crime Pattern Analysis, Visualization and Prediction using Data Mining." *IJARIIE* 1, no. 4 (2015): 681-686.

14. Thongtae, P., and S. Srisuk. "An analysis of data mining applications in crime domain." 2008 IEEE 8th International Conference on Computer and Information Technology Workshops. IEEE, 2008.

15. Sathyadevan, Shiju, and Surya Gangadharan. "Crime analysis and prediction using data mining." 2014 First International Conference on Networks & Soft Computing (ICNSC2014). IEEE, 2014.

16. McClendon, Lawrence, and Natarajan Meghanathan. "Using machine learning algorithms to analyze crime data." Machine Learning and Applications: An International Journal (MLAIJ) 2.1 (2015): 1-12.

17. Malathi, A., and S. Santhosh Baboo. "Enhanced algorithms to identify change in crime patterns." International Journal of Combinatorial Optimization Problems and Informatics 2.3 (2011): 32-38.

18. Nath, Shyam Varan. "Crime pattern detection using data mining." 2006 IEEE/WIC/ACM International Conference on Web Intelligence and Intelligent Agent Technology Workshops. IEEE, 2006.

19. Yu, Chung-Hsien, et al. "Crime forecasting using data mining techniques." 2011 IEEE 11th international conference on data mining workshops. IEEE, 2011.

20. Pande, Vineet, Viraj Samant, and Sindhu Nair. "Crime Detection using Data Mining." International Journal of Engineering Research & Technology (IJERT) http://www. ijert. org ISSN (2016): 2278-0181.

21. Almanie, Tahani, Rsha Mirza, and Elizabeth Lor. "Crime prediction based on crime types and using spatial and temporal criminal hotspots." *arXiv preprint arXiv:1508.02050*(2015).

22. Sivaranjani, S., S. Sivakumari, and M. Aasha. "Crime prediction and forecasting in Tamilnadu using clustering approaches." 2016 International Conference on Emerging Technological Trends (ICETT). IEEE, 2016.

barForecasting of future crime using Data Mining technique    13

23. Chen, Peng, Hongyong Yuan, and Xueming Shu. "Forecasting crime using the arima model." 2008 Fifth International Conference on Fuzzy Systems and Knowledge Discovery. Vol. 5. IEEE, 2008.
24. Zubi, Zakaria Suliman, and Ayman Altaher Mahmmud. "Using data mining techniques to analyze crime patterns in the libyan national crime data." Recent advances in image, audio and signal processing 8 (2014): 79-85.
25. Agarwal, Anisha, et al. "Application for analysis and prediction of crime data using data mining." International Journal of Advanced Computational Engineering and Networking, ISSN (2016): 2320-2106.
26. Sharma, Anshu, and Raman Kumar. "Analysis and design of an algorithm using data mining techniques for matching and predicting crime." International Journal of Computer Science and Technology, IJCST 4.2 (2013): 670-674.
27. Mookiah, Lenin, William Eberle, and Ambareen Siraj. "Survey of crime analysis and prediction." The Twenty-Eighth International Flairs Conference. 2015.
28. Chauhan, Tirthraj, and Rajanikanth Aluvalu. "Using big data analytics for developing crime predictive model." RK University's First International Conference on Research & Entrepreneurship. 2016.
29. Bharti, Sushant, and Ashutosh Mishra. "Prediction of Future Possible Offender's Network and Role of Offenders." 2015 Fifth International Conference on Advances in Computing and Communications (ICACC). IEEE, 2015.
30. Awal, M. A., Rabbi, J., Hossain, S. I., & Hashem, M. M. A. (2016, May). Using linear regression to forecast future trends in crime of Bangladesh. In *2016 5th International Conference on Informatics, Electronics and Vision (ICIEV)* (pp. 333-338). IEEE.
31. Chandrasekar, Addarsh, Abhilash Sunder Raj, and Poorna Kumar. "Crime Prediction and Classification in San Francisco City." *URL http://cs229. stanford. edu/proj2015/228 {\_} report. pdf* (2015).
32. Saeed, U., Sarim, M., Usmani, A., Mukhtar, A., Shaikh, A., & Raffat, S. (2015). Application of machine learning algorithms in crime classification and classification rule mining. *Research Journal of Recent Sciences*, *4*(3), 106-114.
33. Chen, Xinyu, Youngwoon Cho, and Suk Young Jang. "Crime prediction using Twitter sentiment and weather." 2015 Systems and Information Engineering Design Symposium. IEEE, 2015.
34. Mitchell, Mark B., Donald E. Brown, and James H. Conklin. "A crime forecasting tool for the web-based crime analysis toolkit." 2007 IEEE Systems and Information Engineering Design Symposium. IEEE, 2007.
35. Ismail, Suzilah, and Nurulhuda Ramli. "Short-term crime forecasting in Kedah." Procedia-Social and Behavioral Sciences 91 (2013): 654-660.
36. Rosser, G. Predictive crime mapping: Arbitrary grids or street networks? Journal of Quantitative Criminology, 2016, 1–26.
37. Nasridinov, A., yong Byun, J., Um, N., and Shin, H. A study on danger pattern prediction using data mining techniques. Advanced Science and Technology Letters, 2014, 93–96.
38. Wang, Tong, et al. "Learning to detect patterns of crime." Joint European conference on machine learning and knowledge discovery in databases. Springer, Berlin, Heidelberg, 2013.

39. Bogomolov, Andrey, Bruno Lepri, Jacopo Staiano, Nuria Oliver, Fabio Pianesi, and Alex Pentland. "Once upon a crime: towards crime prediction from demographics and mobile data." In *Proceedings of the 16th international conference on multimodal interaction*, pp. 427-434. ACM, 2014.
40. Cavadas, B., Branco, P. and Pereira, S., 2015, September. Crime Prediction Using Regression and Resources Optimization. In Portuguese Conference on Artificial Intelligence (pp. 513-524). Springer, Cham.
41. Grover, Vikas, Richard Adderley, and Max Bramer. "Review of current crime prediction techniques." International Conference on Innovative Techniques and Applications of Artificial Intelligence. Springer, London, 2006
42. Rani, Anchal, and S. Rajasree. "Crime trend analysis and prediction using mahanolobis distance and dynamic time warping technique." International Journal of Computer Science & Information Technologies 5.3 (2014): 4131-4135.
43. Gorr, Wilpen, and YongJei Lee. "Early Warning System for Crime Hot Spots." (2013).
44. Hanslmaier, Michael, et al. "Forecasting crime in Germany in times of demographic change." European Journal on Criminal Policy and Research 21.4 (2015): 591-610.
45. Renushe, Hanmant N., et al. "Short term crime forecasting for prevention of crimes: A study of Satara district." Int. J. Comp. Tech. Appl 2.3 (2012): 608-611.
46. Saoumya, Anurag Singh Baghel. "A predictive model for mapping crime using big data analytics." Int J Res Eng Technol 4.4 (2015).
47. Uk crime and policing in England, Wales and Northern Ireland. (2017). https://data.police.uk/ (accessed on August 14, 2017).